# Atomic-scale composition of the ternary III-V semiconductor (Al,Ga)Sb visualized by cross-sectional scanning tunneling microscopy

Hitesh Kumar,[1,a] Vladimir Kaganer,[1] Zijin Lei,[2] Rüdiger Schott,[2] Werner Wegscheider,[2] and Stefan Fölsch[1,a]

[1]*Paul-Drude-Institut für Festkörperelektronik, Hausvogteiplatz 5-7, Leibniz-Institut im Forschungsverbund Berlin e. V., 10117 Berlin, Germany*

[2]*Laboratory for Solid State Physics, ETH Zürich, 8093 Zürich, Switzerland*

**ABSTRACT.** Cross-sectional scanning tunneling microscopy at 5 K is used to investigate cation mixing in (Al,Ga)Sb layers grown by molecular beam epitaxy, via direct atom counting at the $(1\bar{1}0)$ and (110) cleavage planes. Electronic contrast between Al and Ga surface cations enables statistical analysis of the metal sublattice along the non-equivalent ⟨110⟩ directions within the zincblende (001) surface and along the [001] growth direction. The cation distribution is found to be random both along the growth direction and within the growth plane, with no evidence of long-range order or anisotropic growth kinetics; notably, the mean numbers of consecutive cations of the same type along the two in-plane directions are equal, possibly due to statistical averaging over randomly distributed subsurface cations. The results are compatible with either strain-mediated interactions during cation incorporation at the growth front or ideal, uncorrelated cation mixing. Overall, the examined (Al,Ga)Sb alloy shows an exceptionally high degree of atomic-level homogeneity.

**Orcid IDs**

| | |
|---|---|
| H.K. | 0009-0000-4449-0839 |
| V.K. | 0000-0001-9146-8503 |
| Z.L. | 0000-0002-1420-320X |
| R.S. | 0000-0002-9320-0313 |
| W.W. | 0000-0002-2138-5558 |
| S.F. | 0000-0002-3336-2644 |

[a] Contact authors: kumar@pdi-berlin.de, foelsch@pdi-berlin.de

## I. INTRODUCTION

III-V semiconductor heterostructures form the basis for carrier dynamics in low dimensions – the operating principle behind numerous applications in electronics, optoelectronics, and quantum technology [1,2]. Ternary alloys such as (Al,Ga)As and (Al,Ga)Sb are often the key components of such heterostructures because they offer tunable bandgaps and band alignments across the entire composition range [3]. Since the lattice matching between their corresponding binary alloys is nearly perfect (GaAs-AlAs ~0.1% mismatch) or reasonably close (GaSb-AlSb ~0.6%), standard solution theory [4,5] predicts a vanishing or neglectable miscibility gap for (Al,Ga)As and (Al,Ga)Sb, respectively. However, this does not guarantee that real samples grown by, for example, molecular beam epitaxy (MBE) are free of inhomogeneities in composition. The latter can be detrimental to device performance, for example, in infrared detectors featuring (Al,Ga)Sb barriers [6]. Moreover, long-range order has been reported to occur in (Al,Ga)As grown under specific conditions [7], which clearly goes beyond standard solution theory. All this shows that microscopic information is crucial for determining and understanding the composition of semiconductor alloys at the atomic level [8].

A widely used method in this context is scanning transmission electron microscopy (STEM) which allows for atom-column-resolved composition mapping [9]. In contrast, cross-sectional scanning tunneling microscopy (XSTM) provides pure surface information and is capable of resolving individual atoms at the cleavage surface of a specimen. This was first demonstrated on (110)-cleaved GaAs [10] and later extended to III-V semiconductor heterostructures [11]. XSTM has been used to quantify a wide range of structural and electronic properties of various semiconductor materials; an overview of this extensive research can be found in Ref. [8].

In this work, we employ XSTM to investigate the composition of MBE-grown (Al,Ga)Sb layers – as part of a (Al,Ga)Sb/GaSb superlattice – and perform a statistical analysis of the cation mixing within the metal sublattice of the III-V alloy by a direct count of individual atoms. Specifically, the $(1\bar{1}0)$ and (110) cleavage planes are examined to track the arrangement of cations along both $\langle 110 \rangle$ directions within the zincblende (001) lattice plane, which are not equivalent in terms of symmetry. The experimental results are discussed in the context of the role of growth kinetics, strain-mediated interactions, and ideal cation mixing.

## II. EXPERIMENTAL METHODS

The superlattice samples in this work were grown in a modified Veeco Instruments Gen II MBE system. First, the Te-doped GaSb (001) semi-insulating substrate was thermally degassed under vacuum. The wafer was then transferred into the growth chamber. During the heating process, an Sb flux was supplied and gradually increased once the substrate temperature exceeded 400 °C. After deoxidizing the GaSb surface under an Sb overpressure at 520 °C, the first 50 nm GaSb buffer layer was grown at the same temperature with a growth rate of 0.2 nm/s, as calibrated by reflection high-energy electron diffraction. Subsequently, the (Al,Ga)Sb/GaSb superlattice was grown continuously at 520 °C until the final GaSb layer, during which the substrate temperature was gradually reduced to 480 °C. This lower temperature was optimized for the GaSb/InAs transition. Finally, a thick InAs cap layer was deposited at 480 °C. Further details about the growth method can be found in Ref. [12].

The MBE-grown samples were cleaved in ultrahigh vacuum at a sample temperature of −120 °C and transferred into a cryogenic STM cooled to a sample base temperature of 5 K. The samples were cleaved to expose either the $(1\bar{1}0)$ or the (110) surface termination. STM topography images were recorded in the constant-current mode. Tunneling bias voltages $V$ refer to the sample with respect to the STM tip. Scanning tunneling spectroscopy (STS) spectra were recorded in variable-$z$ mode [13], where an offset $\Delta s(V)$ is applied to the tip-sample separation $s$ which varies linearly with $|V|$. The exponential increase in conductance arising from this variation in $s$ was then normalized by multiplying the data by a factor of $e^{2\kappa\Delta s(V)}$, with $\kappa \cong 1$ Å$^{-1}$ the decay constant. This mode expands the dynamic range of the measurement, allowing for an accurate determination of band-edge positions. Spatial maps of the differential tunnel conductance were recorded at a given bias and at constant tip height. All STS measurements were done using a lock-in amplifier at a peak-to-peak modulation voltage of 20 mV and a modulation frequency of 675 Hz. We used electrochemically etched tungsten tips cleaned by Ne ion sputtering and electron beam heating. A final tip conditioning was accomplished by *in situ* voltage pulsing on the InAs layer area of the cleaved heterostructure; initially developed for tip forming on the InAs(111)A surface [14], this procedure works essentially in the same way on (110)-terminated InAs.

## III. RESULTS AND DISCUSSION

### A. Crystallographic orientation and overall structure of the superlattice

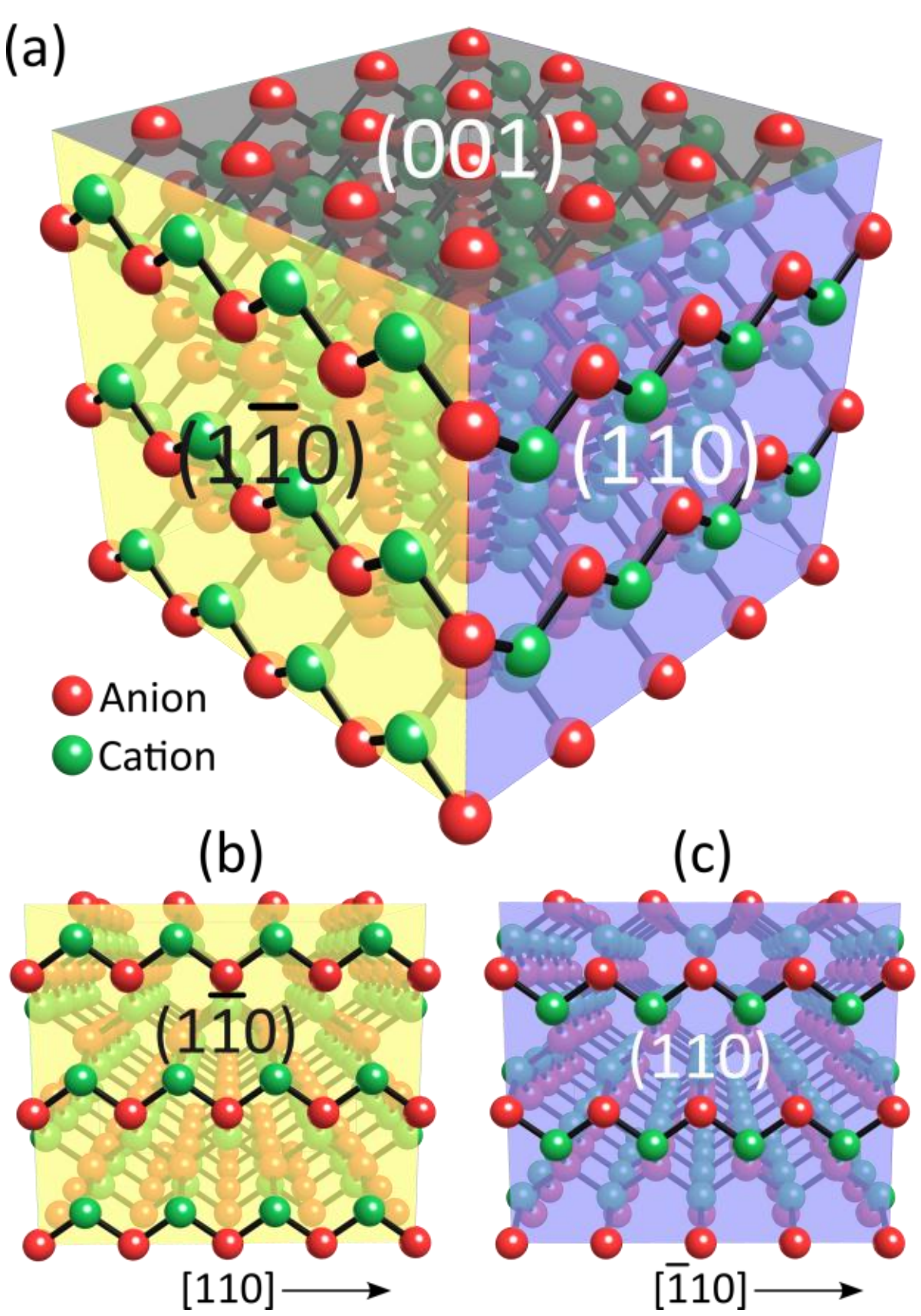

*Fig. 1. (a) Perspective view of the growth (001) plane (gray) relative to the cleavage planes having* (1$\bar{1}$0) *[yellow] and (110) orientation (blue). The arrangement and bonding configuration of the surface anions (red spheres) and cations (green spheres) on the cleavage planes is indicated. Lower panels: the cleavage planes are shown in top view in* (1$\bar{1}$0) *[panel (b)] and (110) cross-section [panel (c)].*

Figure 1(a) illustrates the relative orientations between the (001) growth plane and the (1$\bar{1}$0) and (110) cleavage planes investigated in this work; the coordinate system is chosen such that the anion in the zincblende unit cell is at (0,0,0) and the cation at (¼,¼,¼). Panels (b) and (c) add top-view representations of the two cleavage planes. Because the (001) plane has $C_{2v}$ symmetry (the zincblende structure lacks an inversion center), the [110] and [$\bar{1}$10] directions of the intersecting lines between the growth plane and the two cleavage planes are nonequivalent. Zigzag-shaped rows of surface anions and cations are aligned along these directions within the cleavage planes. The difference in the respective atomic arrangement within the rows can be distinguished by the chemical contrast between the anions and cations in STM imaging [15]. For the cleaved heterostructure investigated here, this difference is also obvious from individual In adatoms transferred from the STM tip to the InAs surface area because those adatoms create an asymmetric band bending arising from their bonding geometry [16].

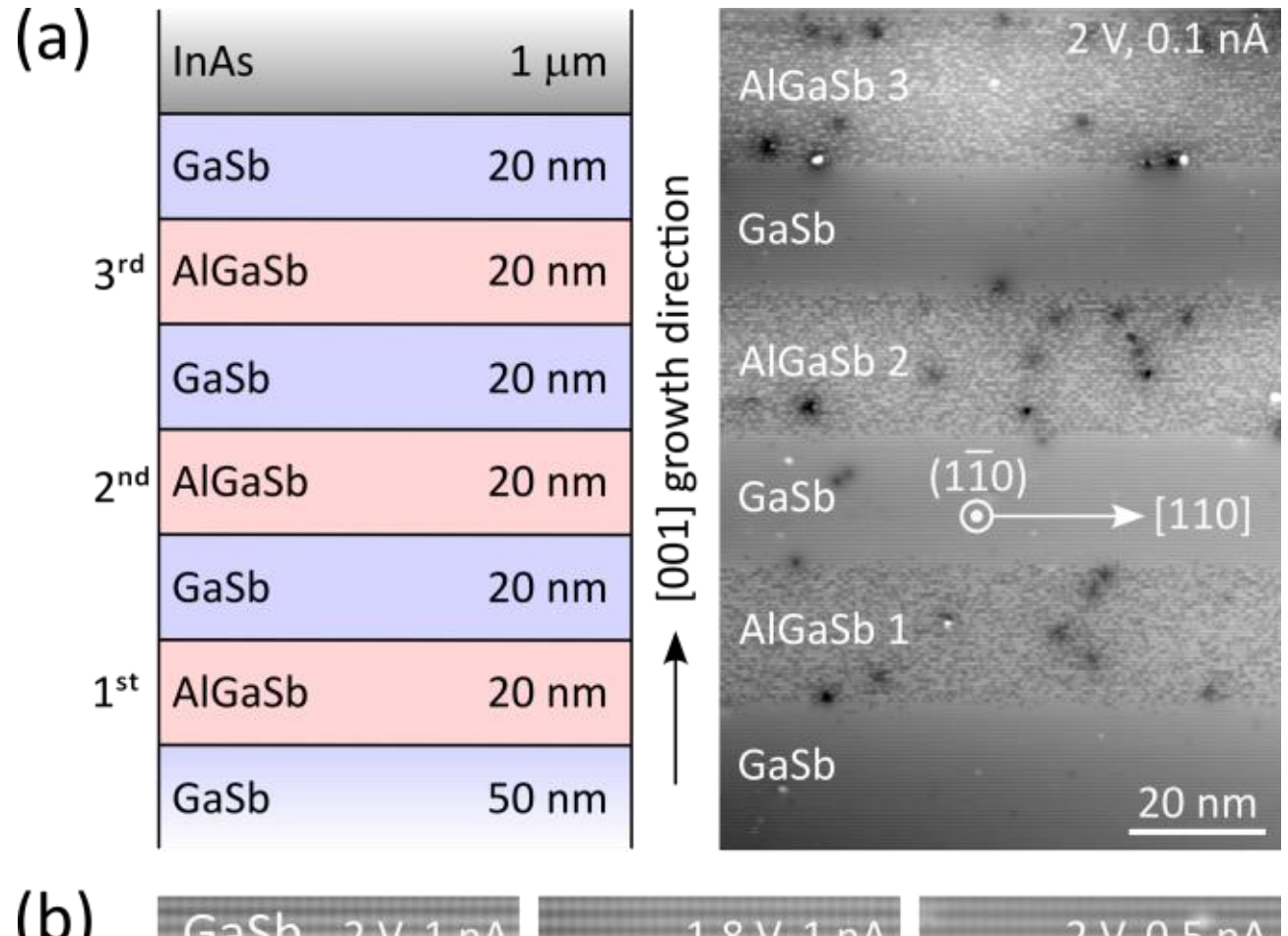

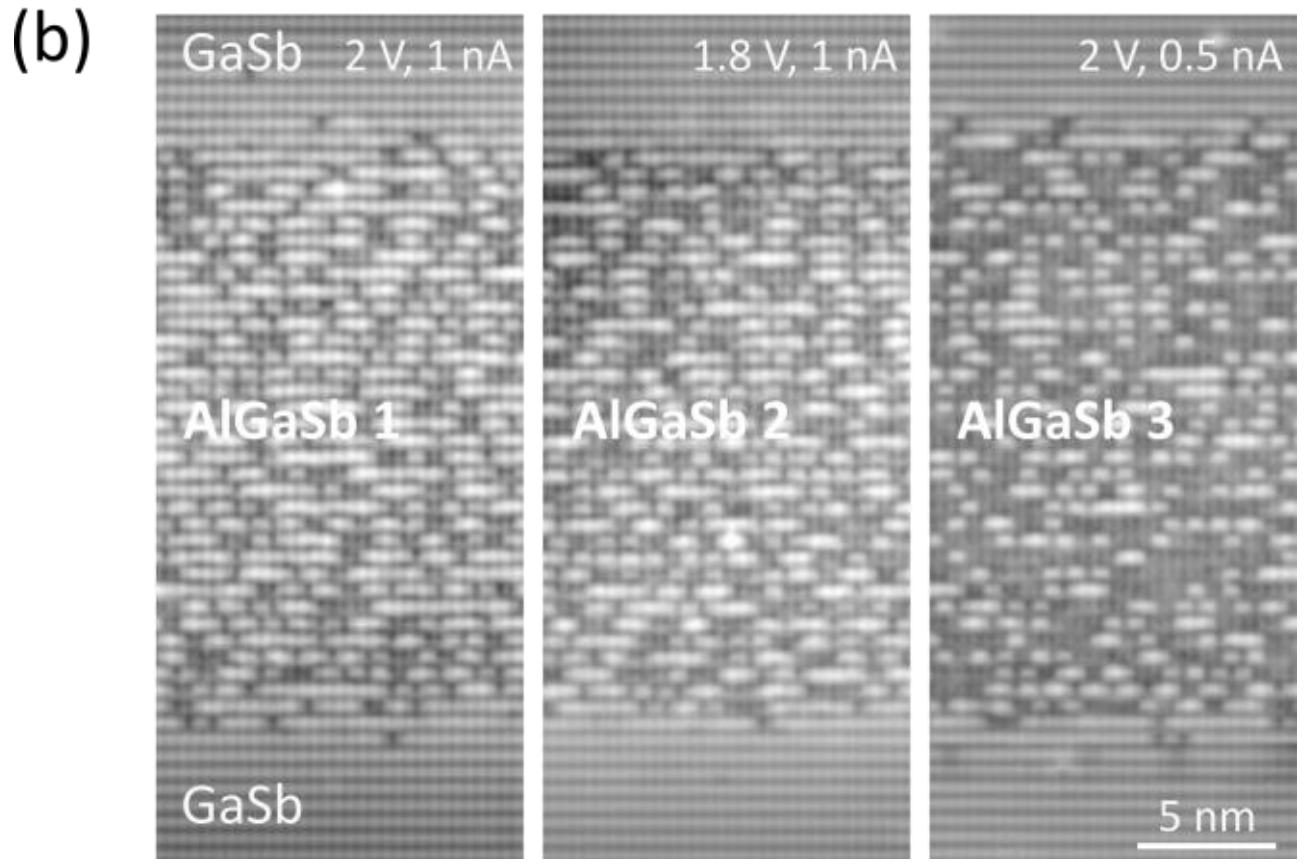

*Fig. 2. (a) Left: structure of the GaSb/(Al,Ga)Sb superlattice grown by MBE on Te-doped GaSb(001) and finished with an InAs capping layer; the Al content is constant in each (Al,Ga)Sb layer and increases from layer 1 to 3. Right: corresponding empty-state STM image of the superlattice in* $(1\bar{1}0)$ *cross-section showing the AlGaSb layers sandwiched between GaSb and numbered along the growth direction. (b) Empty-state topography images of the three (Al,Ga)Sb layers at atomic resolution with chemical contrast between the surface cations Al (smaller apparent height) and Ga (larger height).*

The diagram in Fig. 2(a) shows the GaSb/(Al,Ga)Sb superlattice in $(1\bar{1}0)$ cross-section with the 20-nm-thick (Al,Ga)Sb layers having different Al concentrations; the layers are numbered along the [001] growth direction and the Al content increases from layer 1 to 3. The corresponding STM image on the right hand-side of Fig. 2(a) shows the superlattice with the three (Al,Ga)Sb layers spaced by GaSb layers. The image was recorded at positive sample bias probing the empty states of the cations at the III-V semiconductor surface [15], here Al and Ga. The dark depressions observed at various locations change their appearance to bright protrusions when imaged at negative sample bias, suggesting that they result from the local band bending caused by negatively charged point defects at or near the surface. Empty-state imaging of the three (Al,Ga)Sb layers at atomic resolution [Fig. 2(b)] reveals chemical contrast between the surface cations Al and Ga, with Ga being several 10 pm taller in apparent height compared to Al. (We note that a similar contrast was reported earlier in an XSTM study of GaSb/AlSb interfaces [17].) Essentially the same topography images are observed for the (110) cleavage plane.

**B. Cation-selective contrast in STM imaging**

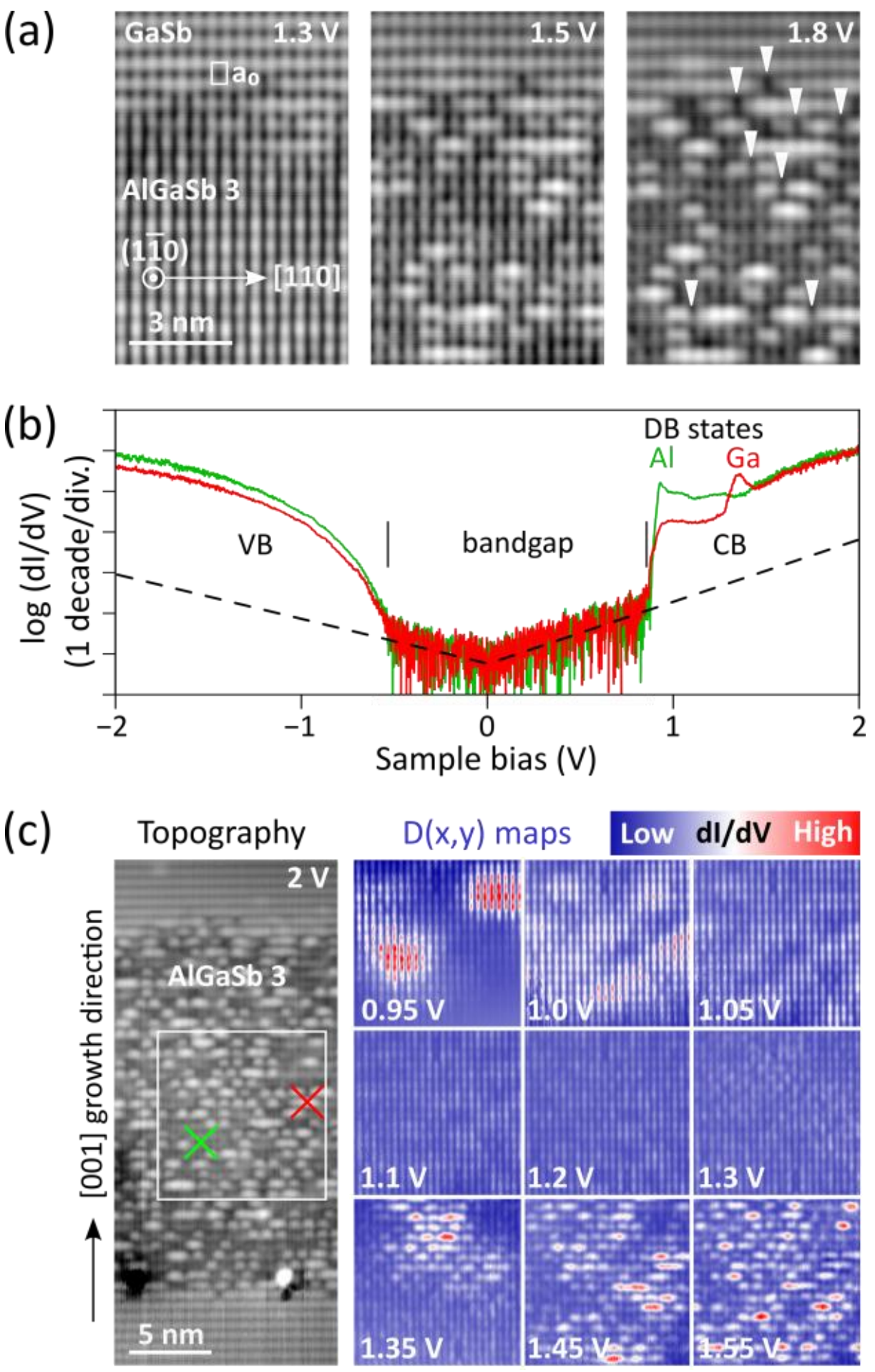

*Fig. 3. (a) Empty-state images recorded at 1 nA and at different positive sample biases; the rectangular surface unit cell is shown ($a_0$ is the cubic lattice constant along [001]), and a defect observed between surface Ga atoms is highlighted by arrows. (b) Conductance spectra recorded with the tip probing Al (green) and Ga aggregates (red) on the third (Al,Ga)Sb layer revealing spectral peaks in the conduction band (CB) region arising from dangling-bond (DB) states of surface Al and Ga. The dashed line is the noise level of the measurement done in variable-z mode. (c) Left: empty-state topography image of the third (Al,Ga)Sb layer, crosses mark the tip positions used in (b). Right: spatial conductance maps D(x,y) recorded within the area highlighted in the topography image, the D(x,y) signal associated with surface Al dominates near the CB minimum while that of surface Ga emerges at 1.35 V.*

We now address the physical origin of the chemical contrast between the two different surface cations. The empty-state images in Fig. 3(a) were recorded on the third (Al,Ga)Sb layer in (1$\bar{1}$0) cross-section interfaced with the subsequent GaSb layer shown in the upper part of the images. It is found that the contrast depends on the actual bias voltage and appears at a bias larger than ~1.3 V, which is an indication of its electronic origin. Arrows in the right hand-side image mark a defect type appearing as a depression between two adjacent Ga cations. This defect occurs within Ga aggregates on (Al,Ga)Sb as well as in the interface region. It is noted that previous work found a defect of similar appearance at GaSb/AlSb interfaces [17]. Although we cannot provide conclusive evidence based on the present data, it is possible that the depression arises from interference effects caused by phase shifts between the dangling-bond orbitals of adjacent surface Ga atoms, leading to a locally reduced tunneling conductance. These phase shifts are apparently resulting from structural inhomogeneities inherent to the (Al,Ga)Sb region and the GaSb/(Al,Ga)Sb interface (the defect does not occur on the bare GaSb regions of the cleaved heterostructure).

The conductance spectra in Fig. 3(b) do indeed show characteristic states that can be attributed to the different surface cations Al and Ga. The spectra were recorded in variable-*z* mode (cf. Section II) with the tip probing Al (green) and Ga (red) aggregates on the third (Al,Ga)Sb layer and reveal distinct conductance peaks just above the conduction-band edge (at ~0.9 V) for Al and at slightly higher bias voltage for Ga. We attribute these peaks to the known dangling-bond states of the surface cations [18,19,20]. The (110) cleavage surface of zincblende III-V semiconductors undergoes a relaxation in which the surface cation (anion) is displaced inward (outward); as a consequence, the cation dangling-bond state rises above the conduction-band minimum whereas the anion state drops below the valence-band maximum, as first demonstrated for GaAs in Ref. [21]. From the spectra in Fig. 3(b), a band gap of ~1.4 eV is obtained which is consistent with the literature value [3] for (Al,Ga)Sb with an Al concentration of ~40% [the Al concentration of the respective (Al,Ga)Sb layers is determined in Sec. III C].

The spatial conductance maps $D(x,y)$ in Fig. 3(c) corroborate the findings discussed in context with the spectra in panel (b): the electronic density of states (DOS) of Al-rich regions occurs right above the conduction-band edge while the DOS of Ga-rich regions emerges at higher energy. At different locations within the scanned area, the respective surface states show up at slightly different energies, which is especially obvious for the Ga-derived state as can be seen in the $D(x,y)$ maps recorded at 1.35 to 1.55 V. This variation is likely caused by electrostatic potential disorder due to residual charged defects – a known effect in semiconductor heterostructures [22,23,24]. The cation aggregation itself may lead to further variations in energy because of spatial confinement effects or fluctuations in the vertical relaxation [21] of surface cations.

### C. Compositional analysis

The chemical contrast between Al and Ga discussed above can be exploited to directly determine the atomic-scale composition of the ternary material. The Al concentration is readily obtained by counting the number of lattice sites with smaller apparent height (Al) compared to those with larger height (Ga). By analyzing forty-two individual images recorded at macroscopically different surface locations on the $(1\bar{1}0)$ and (110) cleavage surfaces and with each image having 95 × 34 lattice sites (along [110]/$[\bar{1}10]$ and [001]), we obtained mean Al concentrations $\langle x \rangle$ of 0.14, 0.26, and 0.46 for the first, second, and third layer, respectively, with a mean standard deviation of $1.7\times10^{-2}$.

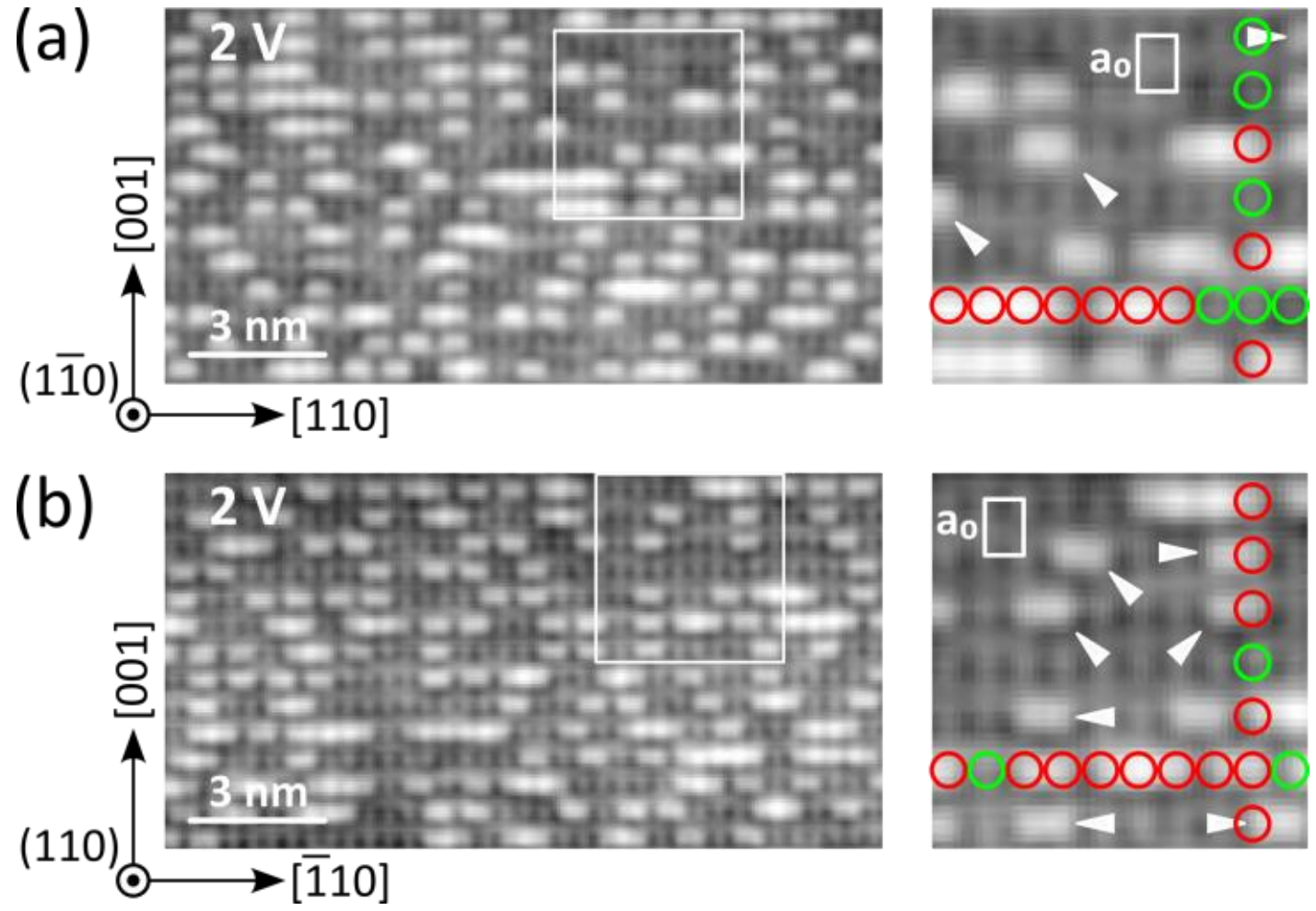

*Fig. 4. (a) Left: empty-state image of the third (Al,Ga)Sb layer in* $(1\bar{1}0)$ *cross-section. Right: close-up view of the area highlighted by the white box in the image on the left, the unit cell of the cation sublattice is indicated. The smallest aggregates observed (marked by arrows) appear as dimers along the [110] in-plane direction. The two different surface cations are highlighted by green (Al) and red (Ga) circles within a selected row along [110] and a column along [001]. (b) Same as (a) but for the third (Al,Ga)Sb layer in (110) cross-section.*

Moreover, we examined the arrangement of the Al and Ga cations along the inequivalent [110] and $[\bar{1}10]$ directions within the growth plane as well as along the [001] growth direction. The procedure is illustrated in Fig. 4, which shows STM images of the third (Al,Ga)Sb layer in the $(1\bar{1}0)$ and (110) cross-sections. The cation arrangement is shown in close-up view on the right hand-side of panels (a) and (b). In the close-up images, the cation sublattice sites occupied by Al (Ga) are marked green (red) along a row within the growth plane and along a column in the direction of growth. The marked sites highlight the two orthogonal directions along which the cation sequence will be tracked row-by-row and column-by-column to extract distributions of the chain lengths $n$ (number of consecutive cations of the same type), as described below. In all the STM images examined, we find that the smallest observable Ga aggregate is two atoms in both $\langle 110 \rangle$ directions within the growth plane [that is, along a zigzag row of surface anions and cations as illustrated in Figs. 1(b) and (c)]. Such aggregates are marked with arrows in the close-up images in Fig. 4. This scenario would imply that the smallest possible Ga aggregate occurring during alloy formation on the (001) growth surface involves four atoms. In contrast, the smallest size observed for surface Al along a $\langle 110 \rangle$ in-plane direction is a single atom, and the same is true for both Al and Ga when tracking the atomic sequence along the [001] growth direction. In the following, we will first consider the scenario of absent Ga monomers, as inferred from the STM images. Second, we will examine the hypothetical case in which Ga monomers are "invisible" to the electronic contrast described above, and discuss its implications.

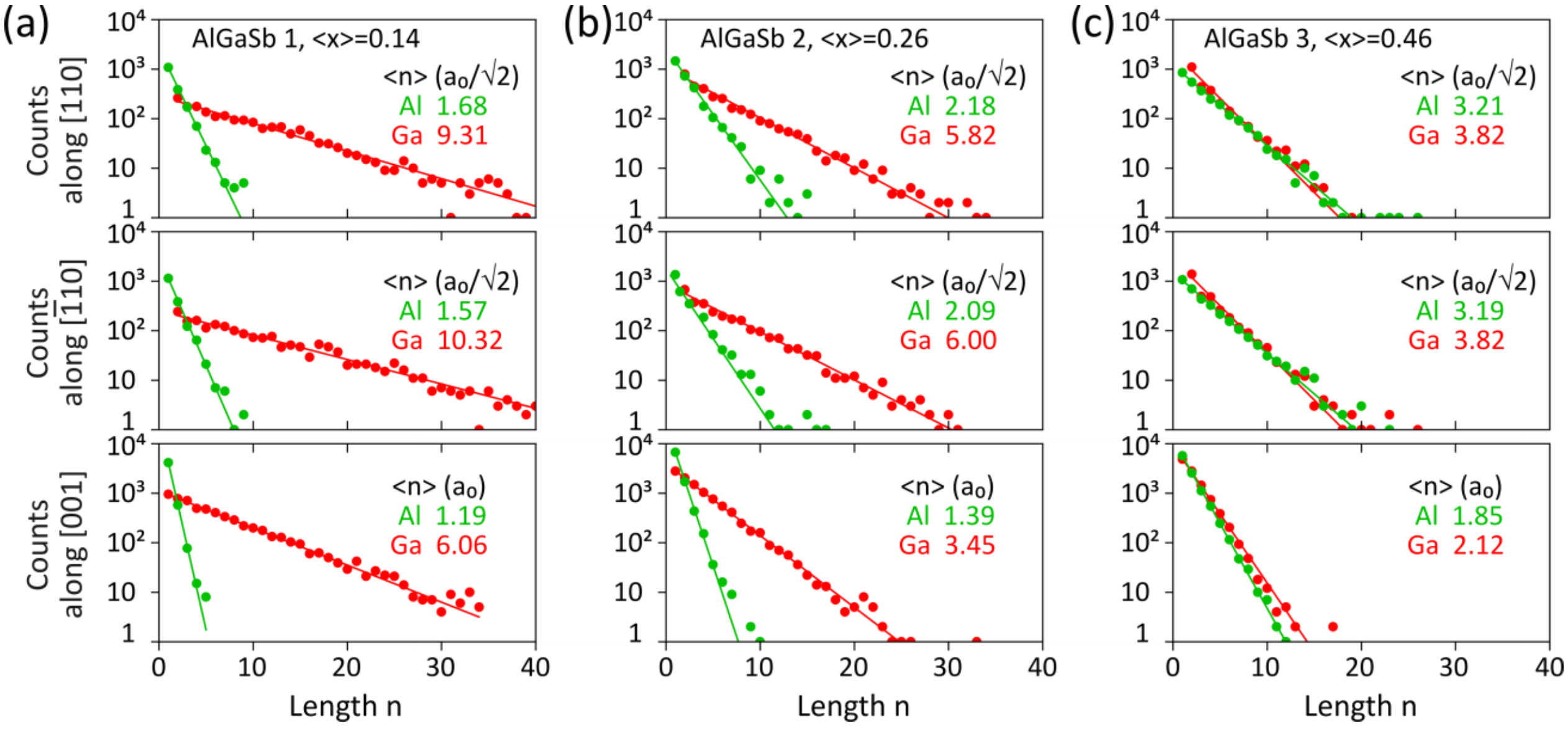

*Fig. 5. (a) Total counts plotted on logarithmic scale versus aggregate length n along [110] (upper plot), [$\bar{1}$10] (center), and [001] (bottom) obtained for the first (Al,Ga)Sb layer; no Ga monomers are observed along the ⟨110⟩ in-plane directions (upper and center plot). Lines in each plot represent geometric distributions with the decay determined by the observed mean length values ⟨n⟩; the same information for the second layer is shown in (b) and for the third layer in (c). The count statistics in panels (a) to (c) was extracted from a total of forty-two individual images each including 95 × 34 cation sublattice sites (along [110]/[$\bar{1}$10] and [001]).*

The three diagrams in Fig. 5(a) summarize the length distributions observed for the first (Al,Ga)Sb layer, $\langle x\rangle$=0.14: plotted are the total counts (on logarithmic scale) versus length $n$ observed along [110] (upper diagram), [$\bar{1}$10] (center), and [001] (bottom). Along the ⟨110⟩ in-plane directions (upper and center diagram), there are no Ga counts for $n$=1, as discussed above. The lines correspond to geometric distributions with their decay given by the experimentally observed mean length values $\langle n\rangle$ listed in the plots. As evident, the agreement with the experimental data points is good, suggesting that the present growth system is well described by random size distributions. The remaining plots in Fig. 5 show the results for the second and the third (Al,Ga)Sb layer, see panels (b) and (c). The chain lengths $n$ follow a random decay at all Al concentrations. Also obvious is that the mean length values $\langle n\rangle$ observed for the ⟨110⟩ in-plane directions are equal in magnitude. Besides this, $\langle n\rangle$ is larger along the two in-plane directions than along the [001] growth direction.

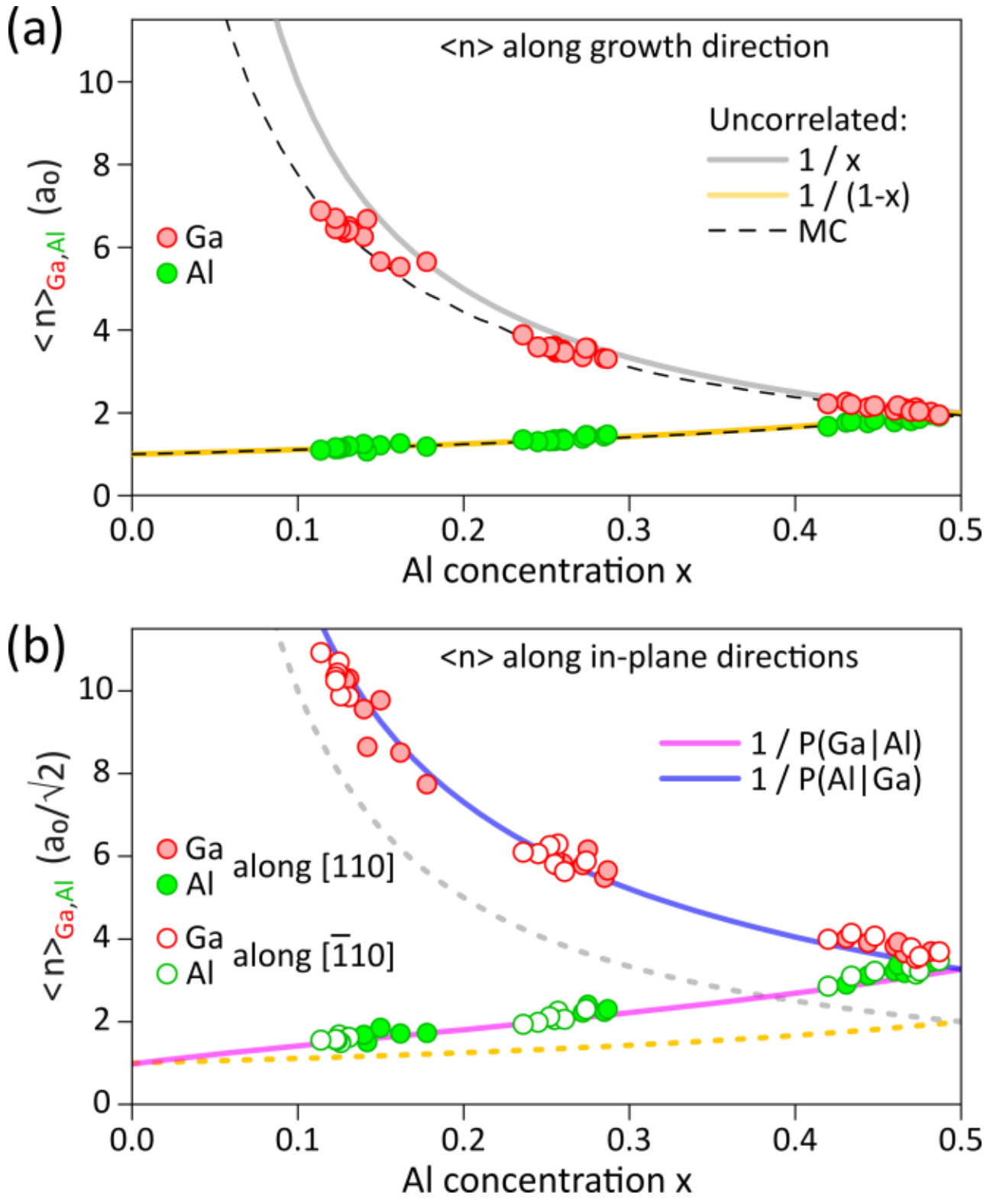

*Fig. 6. (a) Experimental mean length ⟨n⟩ of Ga (red dots) and Al aggregates (green dots) observed along the [001] growth direction plotted versus Al concentration x; each data point was extracted from a single STM image with 95 × 34 sublattice sites (along [110]/[$\bar{1}$10] and [001]). Curves in yellow and gray show the mean length expected for a random and uncorrelated distribution; the black dashed line is a Monte Carlo simulation accounting for the finite layer dimension (thickness) along the growth direction. (b) Mean length ⟨n⟩ observed along the [110] (filled dots) and [$\bar{1}$10] direction (empty dots) within the growth plane, revealing an increase in mean aggregate length as compared to the random and uncorrelated case. Blue and red curves are a fit to a two-state Markov chain model (see main text for details).*

We will now turn to a closer analysis of the data in Fig. 5 and start with the behavior observed along the [001] growth direction which lies in both of the examined cleavage planes. The observed mean length $\langle n \rangle$ can be described within a two-state model in one dimension (1D): let us consider a chain consisting of two types of atoms and assume that the site occupancy is random with no correlation between adjacent sites. The probability $x$ of finding a specific type of atom on a given site is then equal to its concentration. Therefore, the probability $P(n)$ of finding a sequence of $n$ identical atoms is proportional to $x^n$; the requirement of normalization yields $P(n) = x^{n-1}(1-x)$, which corresponds to a geometric distribution with $\langle n \rangle = 1/(1-x)$.

In the diagram in Fig. 6(a), the mean lengths $\langle n \rangle_{\mathrm{Al}}$ (green symbols) and $\langle n \rangle_{\mathrm{Ga}}$ (red symbols) observed along the [001] growth direction are plotted versus the Al concentration $x$. Each data point was obtained from an individual STM image including 95 × 34 lattice sites recorded on the (1$\bar{1}$0) and the (110) cleavage surface, respectively. The random and uncorrelated case described above gives $\langle n \rangle_{Al} = 1/(1-x)$ and $\langle n \rangle_{Ga} = 1/x$, noting that the Ga concentration is $(1-x)$. This prediction is depicted as yellow and gray curves in the plot, yielding fairly good agreement with the data points. The model above implies an infinite chain whereas the counting range along the growth direction is limited to 34 lattice sites because of the finite layer thickness. The resulting finite-size effect is particularly evident at larger mean lengths, with the data points for Ga at smaller *x* lying below the prediction. To account for the finite size, we performed Monte Carlo simulations of a two-state chain with 34 sites and a random, uncorrelated occupancy. The dashed lines in Fig. 6(a) show the average of $10^4$ individual simulations and agree very well with the experimental data.

Figure 6(b) shows the corresponding data points obtained from the cation counting along the ⟨110⟩ in-plane directions. The trend is similar as before, though with larger values of the mean lengths $\langle n \rangle_{\mathrm{Al}}$ and $\langle n \rangle_{\mathrm{Ga}}$. A possible cause of the increased mean length is the presence of strain-mediated interactions resulting from the difference in the bond lengths of Al-Sb and Ga-Sb [3]. Such interactions make it energetically less favorable to have different cations on adjacent sublattice sites, thereby promoting the formation of sequences of identical cations, consistent with the data in Fig. 6(b). This situation can be mimicked by a two-state Markov-chain model described in Appendix A. The model takes into account transition probabilities between adjacent sites rather than the "on-site" probabilities, as they were assumed in the random and uncorrelated case described above. Although the two-state Markov chain is a 1D model and thus a crude simplification of the present problem, we use it here to illustrate the general trend: the blue and pink curves in Fig. 6(b) were fitted to the experimental data under the condition that it is energetically less favorable for adjacent sites to contain different atoms rather than identical ones (see Appendix A for the details).

The experimental data in Fig. 6(b) further reveal that the mean lengths observed along the [110] and $[\bar{1}10]$ directions are essentially the same. This is an unexpected result because these two in-plane directions are inequivalent in terms of symmetry as discussed in connection with Fig. 1. This clearly demonstrates that the cation mixing investigated here is not influenced by effects of anisotropic growth kinetics [25]. The fact that there is no difference in the mean chain lengths along the inequivalent in-plane directions could be due to statistical averaging over the randomly distributed *subsurface* cations, which leads to a cumulative strain field at the growth front that effectively restores isotropy. Such a scenario would be consistent with the general principle of statistical physics, according to which the symmetry of a statistical quantity can be higher than the symmetry of any microscopic configuration of the system [26].

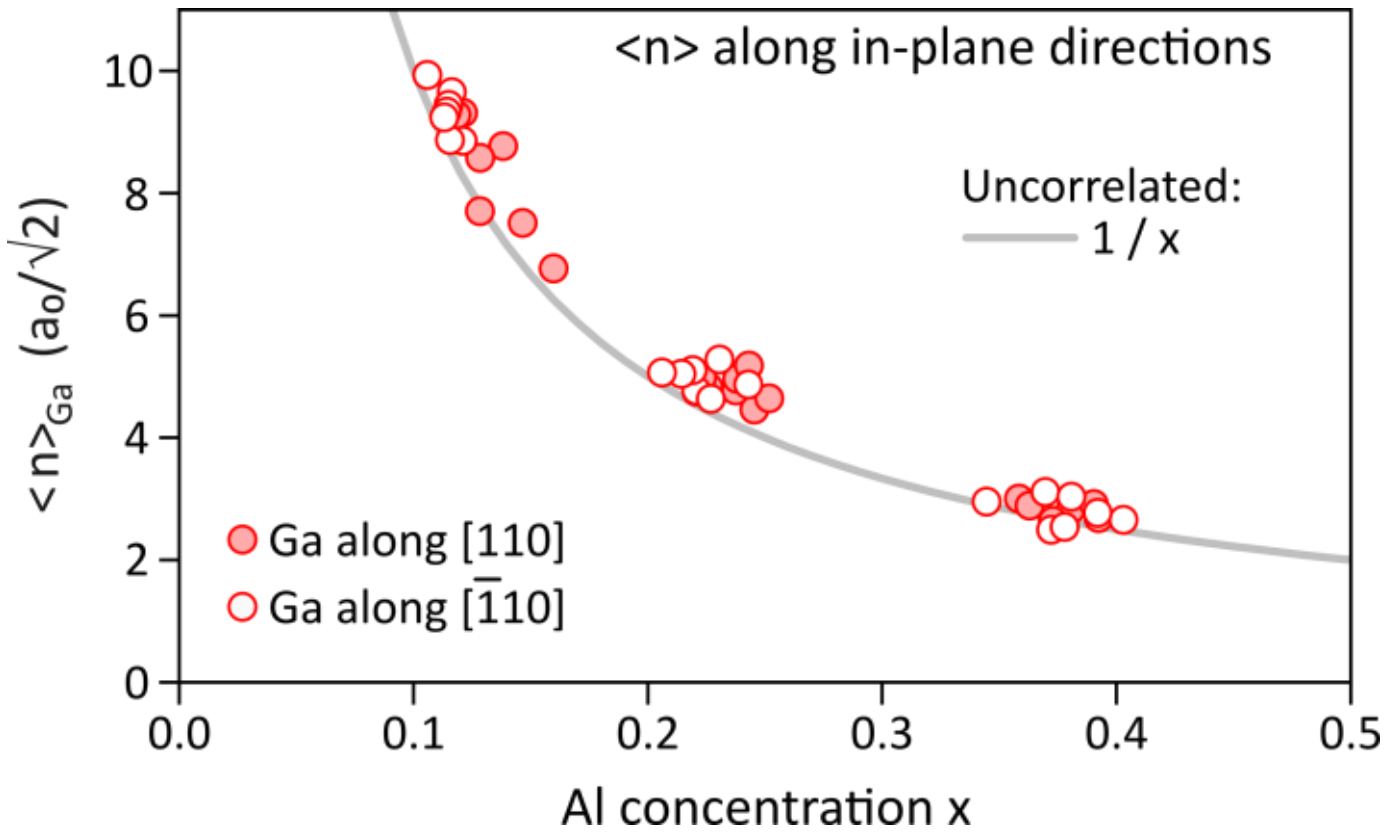

*Fig. 7. Mean length ⟨n⟩ observed along the [110] (filled dots) and [$\bar{1}$10] direction (empty dots) within the growth plane plotted versus Al concentration x, assuming that monomers of Ga surface cations do exist but do not create a chemical contrast (see main text for details). Again, each data point was extracted from a single STM image with 95 × 34 sublattice sites (forty-two images in total). The gray curve shows the mean length expected for a random and uncorrelated distribution.*

Finally, we address the speculative, but not impossible case that single Ga cations at the cleavage surface do not produce the chemical contrast described in Sec. III A and III B. This hypothesis implies that at least two Ga dangling-bond orbitals located at adjacent sites along a surface cation row are required to create a measurable difference in tunneling conductance as compared to Al dangling-bond orbitals. To account for the invisible Ga monomers, we take advantage of the fact that the observed length distributions are well described by geometric distributions and extrapolate the corresponding fraction of monomers. If these are included in the count statistics, both the mean length $\langle n \rangle$ and the Al concentration are decreased. Based on this, we obtain $\langle x \rangle$ values of 0.12, 0.23, and 0.38 for the first, second, and third layer, respectively, and the corresponding $\langle n \rangle$ values plotted in Fig. 7. The result for the mean length of Ga cations is very close to an ideal cation mixing also within the growth plane because the data agree quite well with the random and uncorrelated two-state model discussed above (gray curve in Fig. 7).

## IV. SUMMARY AND CONCLUSIONS

We have presented a detailed compositional analysis of (Al,Ga)Sb layers grown by MBE, based on direct atom counting at the cleavage surface of the ternary III-V alloy. The distinction between Al and Ga atoms is made possible by an electronic contrast in STM imaging, which is presumably due to tunneling via dangling-bond states of the respective surface cations. The counting statistics show that the distribution of cations is random both along the direction of growth and within the plane of growth. There is no indication of long-range order, as it was found earlier for (Al,Ga)As [7] and other III-V alloys [27]. Furthermore, the cation distribution shows no signs of anisotropic growth kinetics [25]. The electronic contrast exploited here allows for two possible interpretations regarding the absence or presence of Ga monomers at the cleavage surface. These two interpretations imply that either strain-mediated interactions are at work when the different cations are incorporated at the growth front, or that there is even an ideal cation mixing – that is, there is no correlation whatsoever between neighboring cations. To decide between these two scenarios, density functional theory could be helpful to elucidate the energetics of the cations at the surface or calculate STM images from first principles [28]. In conclusion, our compositional analysis shows that the (Al,Ga)Sb alloy fabricated and investigated here exhibits an exceptionally high degree of atomic-level homogeneity.

**ACKNOWLEDGEMENTS**

H.K. and S.F. acknowledge funding by the Deutsche Forschungsgemeinschaft (DFG, German Research Foundation) under Grant No. 437494632.

## APPENDIX A: Two-state Markov-chain model

Let us consider a Markov chain of atoms of two kinds, which we call $A$ and $B$ for brevity. The transition probabilities $P(Y|X)$ of having the next atom $Y$ given the current atom is $X$, satisfy the following set of equations. The normalization conditions are

$$P(A|A) + P(B|A) = 1,\ \ P(B|B) + P(A|B) = 1, \tag{1a,b}$$

and the detailed balance (stationarity) condition is

$$x_A P(B|A) = x_B P(A|B), \tag{2}$$

where $x_A$ and $x_B$ are concentrations of $A$ and $B$ atoms ($x_A + x_B$=1). A pair reaction in quasi-chemical equilibrium AB + BA ↔ AA + BB gives a pair-level Boltzmann condition

$$\frac{P(A|B)P(B|A)}{P(A|A)P(B|B)} = e^{-2\Delta\varepsilon/k_B T}, \tag{3}$$

where $\Delta\varepsilon = \varepsilon_{AB} - (\varepsilon_{AA} + \varepsilon_{BB})/2$ is the ordering energy (here $\varepsilon_{XY}$ is the energy of a bond between $X$ and $Y$ atoms), $k_B$ is the Boltzmann constant, and $T$ is the temperature. Equation (3) can be derived as follows. The Markov chain transition probabilities satisfy $P(X|Y) \propto \exp(-\varepsilon_{XY}/k_B T)$, or after normalization by the two-level partition function

$$P(X|Y) = e^{-\varepsilon_{XY}/k_B T} / \left(e^{-\varepsilon_{YA}/k_B T} + e^{-\varepsilon_{YB}/k_B T}\right). \quad (4)$$

Writing the products of the probabilities, we arrive at Eq. (3). The solution of the system (1)-(3) is

$$P(B|A) = \frac{-w + \sqrt{w^2 + 4c_A c_B w(1-w)}}{2c_A(1-w)}, \quad (5)$$

where $w = \exp(-2\Delta\varepsilon/k_B\, T)$; permutation of *A* and *B* in Eq. (5) yields the transition probability $P(A|B)$.

For a comparison with experiment, we are interested in the probability of having *n* consecutive units of identical atoms, for example of type *A*. To build up such a sequence starting from state *A* (a type *A* atom), it requires *n*–1 transitions from *A* to *A* and finally the *n*-th transition from *A* to *B*. The probability is therefore

$$P_A(n) = P(A|A)^{n-1} P(B|A) = P(A|A)^{n-1}[1 - P(A|A)]. \quad (6)$$

This is a geometric distribution with the mean length

$$\langle n \rangle_A = 1/P(B|A) \quad (7a)$$

for type *A* atoms, while an analogous consideration for type *B* atoms yields

$$\langle n \rangle_B = 1/P(A|B). \quad (7b)$$

To compare the model with the experimental data in Fig. 6(b), we set *A*≡Al, *B*≡Ga, and $x_A \equiv x$. Fitting the experimental data to Eqs. (5) and (7a,b) yields *w*=0.19, which is equivalent to an ordering energy $\Delta\varepsilon$ of ~60 meV. This numerical value has no quantitative significance, because the 1D model used here neglects the two-dimensional character of the real system. Of significance is, however, that $\Delta\varepsilon$ is positive (and *w*<1), since this corresponds to the case that different atoms on neighboring sites are energetically less favorable than identical ones.